\documentclass[10pt,twocolumn]{IEEEtran}
\usepackage{cite}

\ifCLASSINFOpdf
\usepackage[pdftex]{graphicx}
\else
\fi
\usepackage{amsmath,amssymb,color, amsthm}
\usepackage{algorithmic}

\usepackage{array}

\usepackage[caption=false,font=footnotesize]{subfig}
\usepackage{float}
\usepackage{url}

\def\bb{{\bf b}}

\def\bI{{\bf I}}

\def\bs{{\bf s}}

\def\by{{\bf y}}

\def\bq{{\bf q}}

\def\bA{{\bf A}}
\def\bB{{\bf B}}

\def\bD{{\bf D}}

\def\bT{{\bf T}}

\newcommand{\bvarphi}{\boldsymbol{\varphi}}

\newcommand{\bDelta}{\boldsymbol{\Delta}}

\newcommand{\bOmega}{\boldsymbol{\Omega}}

\newcommand{\bbeta}{\boldsymbol{\beta}}
\newcommand{\bomega}{\boldsymbol{\omega}}
\newcommand{\balpha}{\boldsymbol{\alpha}}

\newcommand{\btheta}{\boldsymbol{\theta}}
\newcommand{\bPhi}{\boldsymbol{\Phi}}
\newcommand{\brho}{\boldsymbol{\rho}}

\def\minwrt[#1]{\underset{#1}{\text{minimize }}}
\def\argminwrt[#1]{\underset{#1}{\text{arg min }}}
\def\maxwrt[#1]{\underset{#1}{\text{maximize }}}
\def\maxemphwrt[#1]{\underset{#1}{\text{\emph{maximize} }}}

\newtheorem*{remark}{Remark}

\floatstyle{ruled}
\newfloat{algorithm}{t}{lop}
\floatname{algorithm}{Algorithm}

\begin{document}
%
\title{Determining Joint Periodicities in Multi-time Data With Sampling Uncertainties
\thanks{D. Svedberg is with the Division of Signals and Systems, Department of Electrical Engineering, Uppsala University, SE-751 05 Uppsala, Sweden (email: david.svedberg@angstrom.uu.se). F. Elvander is with the Stadius Center for Dynamical Systems, Signal Processing and Data Analytics, KU Leuven, 3001 Leuven, Belgium (email: filip.elvander@esat.kuleuven.be). A. Jakobsson is with the Division of Mathematical Statistics, Centre for Mathematical Sciences, Lund University, SE-221 00 Lund, Sweden (email: aj@maths.lth.se)}}
%
%
%

\author{David~Svedberg,
        Filip~Elvander,
        and~Andreas~Jakobsson,~\IEEEmembership{Senior Member,~IEEE}
}

\maketitle

\begin{abstract}
In this work, we introduce a novel approach for determining a joint sparse spectrum from several non-uniformly sampled data sets, where each data set is assumed to have its own, possibly disjoint, and only partially known, sampling times. The potential of the proposed approach is illustrated using a spectral estimation problem in paleoclimatology. In this problem, each data point derives from a separate ice core measurement, resulting in that even though all measurements reflect the same periodicities, the sampling times and phases differ among the data sets. In addition, sampling times are only approximately known. The resulting joint estimate exploiting all available data is formulated using a sparse reconstruction framework allowing for a reliable and robust estimate of the underlying periodicities. The corresponding misspecified Cramér-Rao lower bound, accounting for the expected sampling uncertainties, is derived and the proposed method is shown to attain the resulting bound when the signal to noise ratio is sufficiently high. The performance of the proposed method is illustrated as compared to other commonly used approaches using both simulated and measured ice core data sets.
\end{abstract}

\begin{IEEEkeywords}
Irregular Sampling, Multi-time, Misspecified Modelling, Paleoclimatology
\end{IEEEkeywords}

%
\IEEEpeerreviewmaketitle

\section{Introduction}
\IEEEPARstart{I}{ce} and sea sediment cores provide an excellent record of, among other things, long-term historic climate variability as well as solar activity via the molecular content trapped in the core \cite{vostok,IAaAC,method,gisp,grip}. By studying periodicities in the time-varying composition of the chemical content of the samples, information that allows for distinguishing between natural and anthropogenic climate change can be extracted \cite{ACH}. 

One such commonly used molecular measure is the ratio $\delta^{18}O$, which is linearly linked to the local temperature, serving as a temperature proxy \cite{IAaAC}. However, the estimation of the frequency content for such data records is complicated by two factors. Firstly, measurements from ice and sea sediment cores constitute an irregularly sampled sequence, as, due to uneven accumulation and sedimentation processes, the samples are not uniformly distributed in time \cite{DOMEF, sedimentationRates}. 
Secondly, the actual sampling times of the measurements are only approximately known. Specifically, the sample time of a particular measurement is determined by using so-called age-depth curves, i.e., a function mapping the depth within the ice/sediment core at which the sample is extracted to its age or sampling time. However, as the functional form, as well as the calibration, of the age-depth curves are only approximate, the actual sampling times are uncertain. Furthermore, the length of the data record for a specific geographic location is limited by the thickness of the ice sheet or sediment layer. However, due to the global effect of astronomical influences, such as, e.g., eccentricity and solar activity, on the long-term climate variability, it is expected that measurements collected at different geographic locations should share similar spectral content in the low-frequency bands, thus opening for the possibility of exploiting multiple data sets to decrease the variability of the obtained spectral estimates \cite{BassinotStack,IAaAC}. Due to measurements being obtained from different locations, with different climate effects, the sampling times for different data sets do, in general, not coincide. Furthermore, even though spectra corresponding to different locations are expected to be similar, there are variations in amplitude and phase, as well as larger dissimilarities in the high frequency content \cite{ImbrieSPEC}. Taken together, these factors complicate inferring the spectral content of the ice/sediment cores using standard methods such as periodogram-based approaches or maximum entropy techniques \cite{burg1975maximum}. Commonly used approaches to address this problem include analyzing multiple normalized and stacked records as a single data set \cite{BassinotStack,ImbrieSPEC}. 

In this work, we formulate the spectral estimation for ice/sediment core data as an inverse problem. Specifically, as it has been indicated that the low-frequency content of such measurements is concentrated to a small set of narrowband components \cite{milankovitch,Imbrie943}, we propose to model the time series as a sinusoidal mixture. Furthermore, as the number of narrowband components is expected to be small, albeit unknown, we utilize a sparse reconstruction framework to estimate the spectral content  \cite{sparse,compressed}. 
In order to allow for several data sets with similar but not identical spectral content, the resulting spectral estimate is formed as a barycenter, i.e., a generalized mean, of the spectra corresponding to the individual time series. Specifically, the spectrum is estimated as the one closest, in a weighted $\ell_2$ sense, to a set of spectra consistent with the respective series of measurements, with a sparsity promoting penalty introduced to exploit the \emph{a priori} knowledge that the expected number of signal components is small. By defining the barycenter in terms of the spectrum, as opposed to in terms of the coefficients of the sinusoidal components, we avoid problems caused by differences in the initial phases for the different spectral components as well as for the used data sets. Furthermore, as is shown, the resulting estimator can be formulated as a convex program, allowing the spectral estimate to be reliably found using standard iterative solvers \cite{cvx1,cvx2}. Similar sparse reconstruction techniques have earlier been used for estimating sinusoidal mixtures in noise \cite{Kato_2012}, chirp signals \cite{chirp}, harmonic structures \cite{GENUSSOV2013390}, as well as in spectroscopic applications \cite{spectroscopy, DYAR201251}. However, to the best of the authors' knowledge, these ideas have not been applied to paleoclimatology data, employing several independent data sets, explicitly accounting for differences in amplitude and phase between them.
Using numerical examples, we demonstrate that the proposed method allows for finding statistically efficient estimates of the frequency content for paleoclimatology data in the ideal case of known sample times, as well as displaying robustness to uncertainty in the sampling. Introducing the misspecified Cramér-Rao lower bound (MCRB) \cite{MCRB2,MCRB} for the problem at hand, we further examine the effects of the sampling uncertainties, illustrating how the achievable performance will depend on this uncertainty. Finally, we examine the estimation performance using both simulated and measured data, clearly showing the preferable performance of the proposed estimator. 

The paper is organized as follows: in the next section, we introduce the problem formulation, after which, in Section \ref{ProposedEstimationApproach}, we introduce the proposed estimator. In Section \ref{VarianceBoundForMisspecifiedData}, we present the MCRB allowing for the assumed sampling uncertainties, after which, in Section \ref{SimulationResults}, we present numerical examples of the discussed estimators, using both simulated and measured data sets. Finally, in Section \ref{Conclusions}, we conclude upon the work. 

\section{Problem Formulation}\label{ProblemFormulation}
Inspired by ice/sediment core data studied in paleoclimatology, we consider data available from $M$ different data sets, for this application typically being attained by analyzing the molecular content of gas trapped in the ice. In this case, the data consists of $\delta^{18}O$ measurements, formed as the ratio of the heavier oxygen isotope ${}^{18}O$ and ${}^{16}O$ as compared to a reference, which will vary with the volume of water in solid vs. liquid form \cite{IAaAC}. From this data, one wants to estimate a common, global, spectrum $\Phi$, describing the frequency content of the climate variation. This spectrum can typically be well-modelled as a collection of point masses, corresponding to representing the signal by a finite sum of sinusoids.
Specifically, the data from core $m$ may be modeled as
\begin{equation}\label{signal}
    s_t^{(m)} = \sum_{k = 1}^K \rho^{(m)}_k \cos\{\omega_k(t+\Delta^{(m)}_t) + \varphi^{(m)}_k\},
\end{equation}
where $\Delta^{(m)}_t$ are the unknown missampling terms, i.e., sampling time errors, at the assumed sampling times $t = t_1^{(m)},\dots,t_{N_m}^{(m)}$, generally being different for each core. The $K$ frequencies, $\omega_k$, are here assumed to be the same for each data set, whereas the amplitudes, $\rho^{(m)}_k$ and phases, $\varphi^{(m)}_k$ are, in general, different for each data set, implying that the data may be detailed using $K+2KM$ unknown parameters.  

The measured signal is modelled as being embedded in additive noise, $\upsilon_t^{(m)}$, such that the observed data is
\begin{displaymath}
    y_t^{(m)} = s_t^{(m)} + \upsilon_t^{(m)}.
\end{displaymath}
Here, we assume that $\upsilon_t^{(m)}$ is normally distributed zero-mean white noise with variance $\sigma_\upsilon^2$, with all observations being independent, such that $y_t^{(m)}\sim \mathcal{N}(s_t^{(m)}, \sigma_\upsilon^2)$. 

Each data set is assumed to be sampled at different sampling instants, with $N_m$ available samples in the $m$-th data set. Let
\begin{displaymath}
    \begin{aligned}
        \bT^{(m)} &\triangleq \begin{bmatrix}t^{(m)}_1 & \dots & t^{(m)}_{N_m} \end{bmatrix}^T\\
        \bDelta_{\bT}^{(m)} &\triangleq \begin{bmatrix}\Delta^{(m)}_{t_1^{(m)}} & \dots & \Delta^{(m)}_{t_{N_m}^{(m)}}\end{bmatrix}^T,
    \end{aligned}
\end{displaymath}
denote the sampling instants and the corresponding missampling terms, respectively, allowing the noise-free and observed signals to be expressed as $(N_m\times 1)$-dimensional vectors
\begin{displaymath}
\begin{aligned}
    \bs^{(m)} &\triangleq \begin{bmatrix}s^{(m)}_{t^{(m)}_1} & \dots & s^{(m)}_{t^{(m)}_{N_m}} \end{bmatrix}^T \\
    \by^{(m)} &\triangleq \begin{bmatrix}y^{(m)}_{t^{(m)}_1} & \dots & y^{(m)}_{t^{(m)}_{N_m}}\end{bmatrix}^T,
\end{aligned}
\end{displaymath}
where the subscript indicate the sampling instants. For notational simplicity, we will, without loss of generality, here assume that the sampling times of all data sets are unique and let $N$ denote the total number of samples available from the $M$ data sets, i.e.,
\begin{displaymath}
    N \triangleq \sum_{m=1}^M N_m.
\end{displaymath}
Combining the $M$ data sets, and defining
\begin{displaymath}
\begin{aligned}
    \bs &\triangleq \begin{bmatrix}(\bs^{(1)})^T & \dots & (\bs^{(M)})^T \end{bmatrix}^T\\
    \by &\triangleq \begin{bmatrix}(\by^{(1)})^T & \dots & (\by^{(M)})^T  \end{bmatrix}^T,
\end{aligned}
\end{displaymath}
we have that the observed $(N\times 1)$-dimensional data vector may be modeled as the multivariate normal distributed vector
\begin{displaymath}
    \by \sim \mathcal{N}(\bs, \sigma_\upsilon^2\bI_{(N\times N)}),
\end{displaymath}
with probability density function (PDF) 
\begin{equation}\label{truemodel}
    f(\by) = \frac{1}{\left(2\pi \sigma_\upsilon^2 \right)^{N/2}}e^{-\frac{1}{2\sigma_\upsilon^2}\sum_{m=1}^M\sum_{t\in\bT^{(m)}}\left(y_t^{(m)}-s_t^{(m)}\right)^2}.
\end{equation}

Thus, the problem of interest is here to determine the $K+2KM$ unknown parameters describing the observations $\by$ while allowing for the sampling uncertainties, $\{\Delta_{t}^{(m)}\}$.

\section{Proposed estimation approach}\label{ProposedEstimationApproach}
The proposed estimator is formed using a sparse reconstruction framework employing a separate wideband dictionary for each of the data sets, as described next. The resulting penalized regression problem exploits that the different data sets share spectral content in some frequency bands, thereby forming an underlying global spectrum, $\Phi$. This spectrum is here represented using the non-negative vector
\begin{displaymath}
    \bPhi \triangleq \begin{bmatrix}\Phi(-\omega_C) & \dots & \Phi(\omega_C) \end{bmatrix}^T \in \mathbb{R}^{2C+1},
\end{displaymath}
where $2C+1$ denotes the number of considered spectral components (which may be different from the number of actual signal components).
The proposed estimator is then formed as the extended penalized regression problem
\begin{equation}\label{cost}
\begin{aligned}
&\minwrt[\{\bbeta^{(m)},\balpha^{(m)}\geq 0\}_{m=1}^M,\bPhi \geq 0 ]{\sum_{m=1}^M||\by^{(m)}-\bD^{(m)}\bbeta^{(m)}||_2^2}\\
&+\sum_{m=1}^M \zeta^{(m)}||\bq\circ (\bPhi-\balpha^{(m)})||_1 + \lambda ||\bPhi||_1 \\
&\text{subject to }
\balpha^{(m)} \geq |\bbeta^{(m)}|,
\end{aligned}
\end{equation}
where $\circ$ denotes element-wise multiplication, and where $D^{(m)}$ denotes the $m$:th (wideband) dictionary, as discussed further below. The user-defined constants $\zeta^{(m)}\geq 0$ reflect the expected correlation between data set $m$ and the rest of the records. Furthermore, 
\begin{displaymath}
    \bbeta^{(m)} \triangleq \begin{bmatrix}\beta^{(m)}(-\omega_C) & \dots & \beta^{(m)}(\omega_C) \end{bmatrix}^T \in \mathbb{C}^{2C+1}
\end{displaymath}
for $m = 1,\dots, M$, and 
\begin{displaymath}
\begin{aligned}
        \balpha^{(m)} &\triangleq \begin{bmatrix} \alpha^{(m)}(-\omega_C) & \dots & \alpha^{(m)}(\omega_C) \end{bmatrix} \in \mathbb{R}^{2C+1} \\
        \bq &\triangleq \begin{bmatrix} q(-\omega_C) & \dots & q^{(m)}(\omega_C) \end{bmatrix} \in \mathbb{R}^{2C+1},
\end{aligned}
\end{displaymath}
where $q(\omega) \in [0,1] \;  \forall \; \omega \in [-\omega_C,\dots, \omega_C]$, representing the individual spectral estimates for the $M$ data sets, as well as a weighting vector, respectively.

The spectral estimate is thus formed as the spectral vector, $\bPhi$, that allows the $M$ data sets to be best represented using the $M$ different dictionaries, $D^{(m)}$, at the assumed sampling times. The second term in the optimization enforces that the estimated global spectrum coincides, at least approximately, with the individual spectral estimates, $\balpha^{(m)}$. The third term is introduced to enforce sparsity on the resulting spectral estimate, whereas the constraint that $\balpha^{(m)}\geq|\bbeta^{(m)}|$ is introduced as the magnitude $|\bbeta^{(m)}|$ is, as illustrated in Appendix \ref{powerloss}, expected to be lower than the global spectral components due to the sampling uncertainties. Thus, the constraint compensate for the loss of power from the missampling, while also ensuring that only differences in magnitude are penalized and thus ignoring differences in phase. By properly weighting these term using the weights $\zeta^{(m)}, m = 1,\dots,M$ and $\lambda$, one can find a trade-off between these properties. Herein, the weights $\zeta^{(m)}$ where, for simplicity, all chosen to be the same, whereas the value of $\lambda$ was chosen such that each of the three terms of the cost function are of similar magnitude, balancing the properties of the estimator. 

Here, the weights in $\bq$ are chosen such that the penalty for a difference in spectral content is only applied to regions where the individual spectra are assumed to be similar. The choice of these regions depends on, among other things, the proximity of the sampling locations, or what time-scale is considered. In a typical example, where one wants to estimate the global spectrum, the shared spectral region would be the low frequency content, such that local, high frequency variations, are ignored. In this example, this would correspond to periods longer than the mean response times of large ice sheets, being roughly 10 thousand years (kyr) \cite{growthRate}.

The $M$ dictionaries, $D^{(m)}$ represent the range of potential frequencies. Here, in order to reduce the resulting computational complexity, we employ the use of iteratively refined wideband dictionaries \cite{WB}. These are formed with the integrated dictionary elements 
\begin{equation}\label{dict}
    [\mathbf{D}]^{(m)}_{n,c} \triangleq \int_{\omega_c^s}^{\omega_c^e}e^{i\omega t^{(m)}_n}d\omega,
\end{equation}
where $\omega_c^s$ and $\omega_c^e$ denotes the start and end frequencies of the $c$:th dictionary element, for $c = 1, \dots, C$, and for time $t^{(m)}_n$. By initially solving \eqref{cost} using a coarse dictionary, the relevant frequency regions may be determined, whereafter the used dictionary may be iteratively refined in order to yield increasing resolution (see also \cite{WB}).

As a sum of norms, the problem in \eqref{cost} is convex and thus can be solved numerically using standard convex solvers, such as CVX \cite{cvx1,cvx2} or SeDuMi \cite{SeDuMi}. 

\begin{remark}
It is worth noting that although the signals of interest here are real-valued, it is still beneficial to make use of a complex-valued dictionary. This as such a dictionary allows for different phases in each core, which occurs, for example, when there are two cores available with a single frequency with the same amplitude but different phases, i.e., $M=2$, $K=1$, and
\begin{align*}
    y^{(m)}_t = \rho \cos(\omega_1 t + \varphi_m), \quad m =1,2.
\end{align*}
Then, the complex-valued amplitude vectors $\bbeta^{(1)}$ and $\bbeta^{(2)}$ will be
\begin{align*}
    \bbeta^{(m)}(\omega_1) &= \frac{\rho}{2}e^{i\varphi_m}, \quad m = 1,2\\
    \bbeta^{(m)}(-\omega_1) &= \frac{\rho}{2}e^{-i\varphi_m}, \quad m = 1,2,
\end{align*}
and the non-negative constrained vectors $\balpha^{(1)} \geq |\bbeta^{(1)}|$,
 and $\balpha^{(2)} \geq |\bbeta^{(2)}|$, i.e.,
\begin{displaymath}
    \balpha^{(m)}(\pm \omega_1) \geq \frac{\rho}{2}, \quad m = 1,2,\\
\end{displaymath}
ensuring the second sum in \eqref{cost} will depend only on the differences in magnitude, enabling the difference in phase to be ignored. Since $\bPhi$ is the magnitude of the common spectrum, the power is given by
\begin{equation}\label{power}
 \hat{p}(\omega_c) = \Phi(-\omega_c)^2+\Phi(\omega_c)^2.
\end{equation}
\end{remark}

The proposed estimator is summarized\footnote{An implementation of the algorithm will be provided on the authors' webpages upon publication.} in Algorithm \ref{freqfind}, and consists of two parts, firstly the wideband zooming procedure described above is employed to find the support of the model. In its final step, a grid-less narrowband search is employed in order to a single frequency in each of the active regions obtained from the earlier zooming step (see also \cite{WB}). The frequency estimates are found as
\begin{equation}\label{NBsearch}
    \hat{\bomega} = \argminwrt[\bOmega^s \leq \bomega \leq \bOmega^e]{\sum_{m=1}^M \left\lVert\by^{(m)}-\left(\bA_{\bomega}^{(m)}\right)^\dagger\by^{(m)}\right\rVert_2^2},
\end{equation}
where $(\cdot)^\dagger$ denotes the matrix pseudoinverse, i.e.,
\begin{displaymath}
    \bA^\dagger = \bA(\bA^T\bA)^{-1}\bA^T,
\end{displaymath}
where $\bA^{(m)}_{\bomega}$ is the $(N_m \times 2C+1)$ real-valued narrowband dictionary for data set $m$, i.e.,
\begin{displaymath}\begin{aligned}
    \bA^{(m)}_{\bomega} = &\left[\begin{matrix}\cos(\omega_1\bT^{(m)}) & \sin(\omega_1\bT^{(m)}) & \dots\end{matrix}\right.\\
    &\left.\begin{matrix}\dots & \cos(\omega_C\bT^{(m)}) & \sin(\omega_C\bT^{(m)})\end{matrix}\right].
    \end{aligned}
\end{displaymath}
This gridless search only results in frequency estimates; if one also wishes to find the amplitudes of the components at these frequencies, these may be found using least squares. A global amplitude estimate can then be formed by taking the mean of the amplitudes of the different data sets. It should be noted that, in general, the amplitudes may differ substantially between cores \cite{IAaAC}. This simplistic approach may be problematic for certain applications where accuracy of the amplitude estimates is of great importance, but here it is only used to provide a visual comparison between different methods.

\begin{algorithm}[!t]
    \caption{The proposed global spectral estimation approach (GLOSA)}
    \begin{algorithmic}[1]
        \STATE Choose the number of zooming steps, $I_z$
        \STATE Choose the number of candidate bands, $C_1$
        \STATE Set the frequency bin $\Delta_1 =\omega_{max}/C_1$, where $\omega_{max}$ is the maximum frequency of interest
        \STATE Set $\bOmega_1^s = \{\omega_c : \omega_c = c\Delta, \text{for } c = 0,\dots,C_1-1\}$, $\bOmega_1^e = \{\omega_c: \omega_c = c\Delta, \text{for } c = 1,\dots,C_1\}$
        \STATE Form the dictionaries $\{\bD_1^{(m)}\}_{m=1}^M$, using \eqref{dict}
        \STATE Solve the minimization problem in \eqref{cost}
        \STATE Calculate $\hat{p}_1(\omega_c), \text{ for } c = 1,\dots,C,$ using \eqref{power}
        \STATE $\boldsymbol{\mathcal{C}}_1 = \{c:\hat{p}_1(\omega_c)>\tau, \text{for } c = 1,\dots C_1\}$
        \STATE $\bOmega^s_{active} = \{\omega_c \in \bOmega^s_1 : c\in \boldsymbol{\mathcal{C}}_1\}$, $\bOmega^e_{active} = \{\omega_c \in \bOmega^e_1: c\in \boldsymbol{\mathcal{C}}_1\}$
        \FOR{$z = 2$ to $I_z$}
            \STATE Choose the number of zoom bands, $C_z$
            \STATE Select frequency bins $\Delta_z = \Delta_{z-1}/C_z$
            \STATE $\bOmega^s_z=\{\bomega_c:\bomega_c=\begin{bmatrix}\omega_c + \Delta_z,\dots,\omega_c+C_z\Delta_z\end{bmatrix}^T,\omega_c\in \bOmega^s_{active}\}$\\
            $\bOmega^e_z = \{\bomega_c : \bomega_c = \begin{bmatrix}\omega_c + \Delta_z,\dots,\omega_c+C_z\Delta_z\end{bmatrix}^T,\omega_c\in \bOmega^e_{active}\}$
            \STATE Form the dictionaries $\{\bD_z^{(m)}\}_{m=1}^M$ using \eqref{dict}
            \STATE Solve \eqref{cost}
            \STATE Calculate $\hat{p}(\omega_c)$ for $c = 1,\dots,\prod_{1}^z C_z $, using \eqref{power}
            \STATE $\boldsymbol{\mathcal{C}}_z = \{c : \hat{p}_z(\omega_c)>\tau, \text{for } c = 1,\dots,\prod_{1}^zC_z\}$
            \STATE $\bOmega^s_{active} = \{\omega_c \in \bOmega^s_z : c\in \boldsymbol{\mathcal{C}}_z\}$, $\bOmega^e_{active} = \{\omega_c \in \bOmega^e_z: c\in \boldsymbol{\mathcal{C}}_z\}$
        \ENDFOR
        \STATE Find $\hat{\bomega}$ by solving the minimization problem in \eqref{NBsearch}
        \STATE Find the amplitude estimates for each core using least squares at $\hat{\omega}$.
        \STATE Form the global spectral estimate as the mean of the amplitudes found in the previous step. 
    \end{algorithmic}
    \label{freqfind}
\end{algorithm}

We now proceed to derive a lower bound on the achievable accuracy of the resulting estimates. 
\section{Variance Bound for Misspecified Data}\label{VarianceBoundForMisspecifiedData}
In this section, we derive a lower bound on the achievable variance for the considered estimation problem using the corresponding misspecified Cramér-Rao lower bound (MCRB). Since the missampling in \eqref{signal} is unknown, and one only has access to the assumed sampling times, the unknown signal may only be approximated. This can be done by disregarding the missampling, introducing the approximate signal model
\begin{equation}\label{approx}
\begin{aligned}
    \mu_t^{(m)}(\btheta) &= \sum_{k=1}^K\bar{\rho}_k^{(m)}\cos(\bar{\omega}_k t+\bar{\varphi}_k^{(m)}),
\end{aligned}
\end{equation}
at the assumed sampling instants $t = t_1^{(m)},\dots, t_{N_m}^{(m)}$, and where $\btheta$ is the collection of model parameters defining the signal, i.e.,
\begin{displaymath}
    \btheta \triangleq \begin{bmatrix}\bar{\bomega}&\bar{\brho}^{(1)}&\hdots&\bar{\brho}^{(M)}&\bar{\bvarphi}^{(1)}&\hdots&\bar{\bvarphi}^{(M)}\end{bmatrix}^T,
\end{displaymath}
where
\begin{displaymath}
    \begin{aligned}
    \bar{\bomega} &\triangleq \begin{bmatrix}\bar{\omega}_1 &\hdots&\bar{\omega}_K\end{bmatrix}\\
    \bar{\brho}^{(m)} &\triangleq \begin{bmatrix}\bar{\rho}_1^{(m)} & \hdots & \bar{\rho}_K^{(m)}\end{bmatrix}, \quad\text{for } m = 1,\dots,M\\
    \bar{\bvarphi}^{(m)} &\triangleq \begin{bmatrix}\bar{\varphi}_1^{(m)} & \hdots & \bar{\varphi}_K^{(M)} \end{bmatrix}, \quad\text{for } m = 1,\dots,M.
    \end{aligned}
\end{displaymath}
It should here be noted that the elements of $\btheta$ are not the same parameters as those describing \eqref{signal} due to the unknown, and thereby disergarded, missampling terms. The measured signal will thus be approximated as the misspecified model in \eqref{approx} being corrupted by an additive zero-mean white Gaussian noise, $\epsilon_t^{(m)}$, with variance $\sigma_{\epsilon}^2$, such that the PDF of the misspecified model, $g_{\btheta}(\by)$, using $\nu_t^{(m)}(\btheta) \triangleq y_t^{(m)}-\mu_t^{(m)}(\btheta)$, may be expressed as
\begin{equation}\label{mismodel}
    g_{\btheta}(\by) = \frac{1}{\left(2\pi\sigma_{\epsilon}^2\right)^{N/2}}e^{-\frac{1}{2\sigma_{\epsilon}^2}\sum_{m=1}^M\sum_{t\in \bT^{(m)}}\left(\nu_t^{(m)}(\btheta)\right)^2}.
\end{equation}
It should be stressed that the model in \eqref{approx} does not take into account the missampling, implying that the resulting model is misspecified. To take this model mismatch into account, we derive the corresponding MCRB (see e.g., \cite{MCRB,MCRB2} for further details on the MCRB), yielding a bound for $\btheta$ under the misspecified model. In order to do so, we first review the notion of the Kullback-Leibler divergence.

\subsection{Relative Entropy - the Kullback-Leibler divergence}
The Kullback-Leibler divergence (KLD), or the relative entropy, between two probability densities $f$ and $g$ is defined as
\begin{displaymath}
    D(f||g) \triangleq \mathbb{E}_{f}\left\{\log\left(\frac{f}{g}\right)\right\},
\end{displaymath}
where $\mathbb{E}_f$ denotes the expectation with respect to the PDF $f$.

Given a certain observation of the missampling errors, the KLD between the true and misspecified model (as derived in Appendix \ref{KLDapp}) is
\begin{equation}\label{KLDshort}
\begin{aligned}
    D(f||g_{\btheta}) 
    &= \frac{1}{2}\Bigg[N\log\left(\frac{\sigma_{\epsilon}^2}{\sigma_\upsilon^2}\right)+ N\left(\frac{\sigma_\upsilon^2}{\sigma_{\epsilon}^2}-1\right)\\&+\left.\frac{1}{\sigma_{\epsilon}^2}\sum_{m=1}^M\sum_{t\in\bT^{(m)}}\left(s_t^{(m)}-\mu_t^{(m)}(\btheta)\right)^2\right].
\end{aligned}
\end{equation}
It is worth noting that the parameter of the misspecified model only affects the final term of the KLD.

\subsection{Pseudo-true parameters}
The next step in forming the MCRB is to determine the corresponding pseudo-true parameters, $\btheta_0$, which are defined as the parameters minimizing the KLD \cite{MCRB,MCRB2}. Examining the final term in \eqref{KLDshort}, one may note that when only the model parameters affect the waveform error, these parameters will coincide with those minimizing the squared $\ell_2$ norm between the approximating signal in \eqref{approx} and the true signal in \eqref{signal}, such that
\begin{displaymath}
    \btheta_0 = \argminwrt[\btheta]{\sum_{m=1}^M\sum_{t\in\bT^{(m)}}(s_t^{(m)}-\mu_t^{(m)}(\btheta))^2},
\end{displaymath}
implying that $\btheta_0$ will minimize the KLD between the distributions of the true and misspecified models.

Since the pseudo-true variance does not affect the waveform error, this may be minimized separately using \eqref{KLDshort}, yielding
\begin{displaymath}
\begin{aligned}
    \hat{\sigma}_{\epsilon}^2 &\triangleq \argminwrt[\sigma_{\epsilon}^2>0]{\frac{1}{2}N\log\left(\frac{\sigma_{\epsilon}^2}{\sigma_\upsilon^2}\right)+N\left(\frac{\sigma_\upsilon^2}{\sigma_{\epsilon}^2}-1\right)}\nonumber\\&\quad+ \frac{1}{\sigma_{\epsilon}^2}\sum_{m=1}^M\sum_{t\in \bT^{(m)}}\left(s_t^{(m)}-\mu^{(m)}_t(\btheta)\right)^2\nonumber \\
    &=\sigma_\upsilon^2+\frac{1}{N}\sum_{m=1}^M\sum_{t\in \bT^{(m)}}\left(s_t^{(m)}-\mu_t^{(m)}(\btheta)\right)^2.
\end{aligned}
\end{displaymath}
Loosely speaking, this means that as the waveform error grows due to the missampling error and the misspecification increases, a higher pseudo-true variance is compensating for the model error. 

\subsection{The Misspecified Cramér-Rao Lower Bound}
Under relatively weak assumptions (see \cite{MCRB,MCRB2} and references therein), the MCRB may be formed using two generalizations of the Fisher information matrix. 

Denoting the true PDF of the observations $f$, the MCRB of any unbiased (w.r.t. the pseudo-true parameters) estimator $\hat{\btheta}$ satisfies \cite{MCRB,MCRB2}
\begin{displaymath}
    \mathbb{E}_f\left\{\left(\hat{\btheta}-\btheta\right)\left(\hat{\btheta}-\btheta\right)^T\right\} \geq \bA_{\btheta_0}^{-1}\bB_{\btheta_0}\bA_{\btheta_0}^{-1},
\end{displaymath}
where (for details on these derivations, see Appendix \ref{FIMA})
\begin{displaymath}
\begin{aligned}
    \bA_{\btheta_0} &\triangleq \mathbb{E}_{f}\{\nabla_{\btheta}^2\ln{g_{\btheta}(\by)}|_{\btheta = \btheta_0}\} \nonumber \\
    &=\frac{1}{\hat{\sigma}_{\epsilon}^2}\sum_{m=1}^M\sum_{t\in\bT^{(m)}}\left[\left(s_t^{(m)}-\mu_t^{(m)}(\btheta_0)\right)\nabla_{\btheta}^2\mu_t^{(m)}(\btheta_0)\right.\nonumber \\
    &\quad\left.-\nabla_{\btheta}\mu_t^{(m)}(\btheta_0)\nabla_{\btheta}^T\mu_t^{(m)}(\btheta_0)\right] \\
    \bB_{\btheta_0} &\triangleq \mathbb{E}_{f}\{\nabla_{\btheta}\ln{g_{\btheta}}(\by)|_{\btheta = \btheta_0}\nabla_{\btheta}^T\ln{g_{\btheta}(\by)}|_{\btheta = \btheta_0}\} \nonumber \\
    &=\frac{\sigma_\upsilon^2}{(\hat{\sigma}_{\epsilon}^2)^2}\sum_{m=1}^M\sum_{t\in \bT^{(m)}} \nabla_{\btheta}\mu_t^{(m)}(\btheta_0)\nabla_{\btheta}^T\mu_t^{(m)}(\btheta_0).
\end{aligned}
\end{displaymath}

Since the misspecified model can be considered a special case of the true model obtained when $\Delta_t^{(m)} = 0 \ \forall t \in \bT^{(m)}$, a lower bound on the mean squared error (MSE) of the frequency estimates may be formed as \cite{MCRB2,MCRB}
\begin{displaymath}
    LB(\btheta_0) = \bA^{-1}_{\btheta_0}\bB_{\btheta_0}\bA^{-1}_{\btheta_0}  + (\tilde{\btheta}-\btheta_0)(\tilde{\btheta}-\btheta_0)^T,
\end{displaymath}
where $\tilde{\btheta}$ denotes the true model parameters in \eqref{signal}.

\section{Simulation results}\label{SimulationResults}
In this section, we evaluate the proposed method on both real and simulated data. To ensure that the simulated data displays the correct characteristics regarding, e.g., sampling irregularities, we employ a randomized sanpling pattern construction, as described next. 
\subsection{Simulating ice core data}\label{Simul}
The sampling pattern used for the simulated data is here based on the Vostok ice core \cite{vostok} and on the Taylor Dome ice core  \cite{taylorAge,taylorAge2}. In order to ensure that each simulation has its own sampling pattern, we simulate each pattern using a mixed model Poisson process with intensity function estimated from the real ice core data using maximum likelihood. 

\begin{figure}[!t]
    \centering
    \includegraphics[width = .49\textwidth]{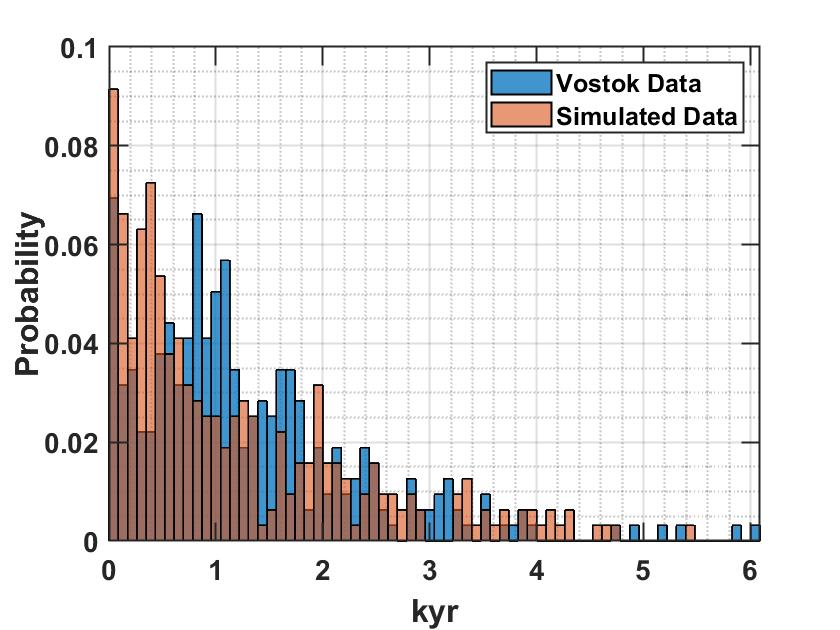}
    \caption{Histogram showing the distribution of sampling distances in the Vostok ice core and from a single simulation. Notice the double peaks of the Vostok data, a characteristic which is emulated using the mixture model in the simulated data, although in this example, the location of the second peak is slightly different.}
    \label{sampdisthist}
\end{figure}

\begin{figure}[!t]
    \centering
    \includegraphics[width = .49\textwidth]{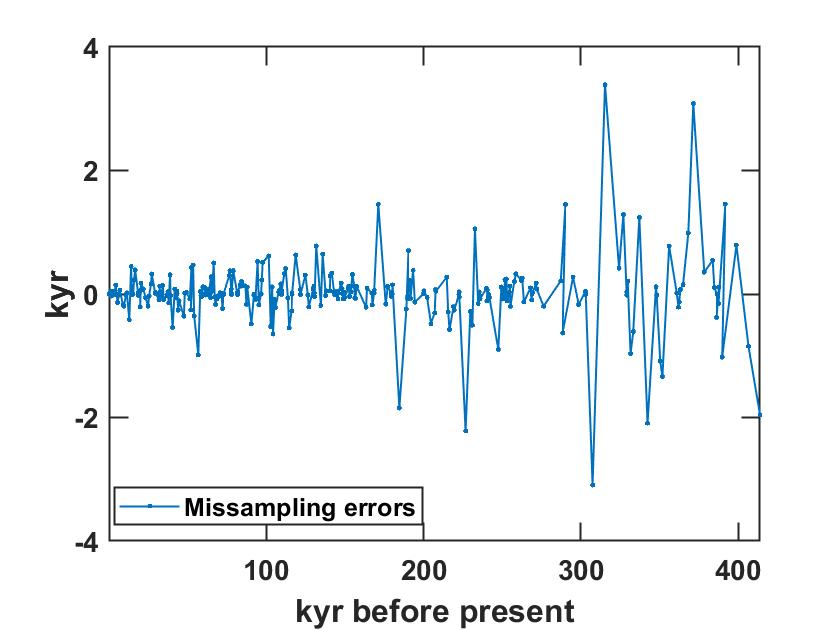}
    \caption{A typical example of how the missampling errors behave over time, the sampling scheme is based on the Vostok ice core. One can easily see that the general trend is that the size of the missampling errors grows with the age of the record. This emulates this characteristic feature from the ice-core data.}
    \label{missampvsT}
\end{figure}

It should be noted that, due to the compression of the ice under its own weight, the sampling distances are increasing with the age of the records. This effect is less pronounced for more recent samples since the ice has had less time to compress. Furthermore, due to the approximating nature of the age vs depth curves, the age uncertainty grows with the age of the records \cite{vostokAge}. To capture this behaviour, the first part of the simulated sampling patterns are modelled using a homogeneous Poisson process which switches to a nonhomogeneous Poisson process at some change point, $T_c$. Let $X(t)$ denote the number of events of a Poisson process with constant intensity $\gamma$ at time $t$, and let $\{T_i\}_{i=1}^{N_m}$ define the sampling instants of the real world data used for simulation. Then, denoting $\Delta T_i \triangleq T_{i+1}-T_i$
\begin{displaymath}
    \mathbb{P}\{X(T_{i+1})-X(T_i) = 1\} = \gamma\Delta T_i e^{-\gamma\Delta T_i},
\end{displaymath}
Using the log-likelihood function, 
\begin{displaymath}
    l(\gamma^c|\{T_i\}_{i=1}^c) = \sum_{i=1}^{c-1}\log(\gamma^c)+\log\Delta T_i - \gamma^c\Delta T_i,
\end{displaymath}
where $T_c\leq T_{N_m}$ denotes the change point, one can find the ML estimate of the intensity $\gamma$ by finding the stationary point of the log-likelihood function,
\begin{displaymath}
    \hat{\gamma}^c_{MLE} = \frac{c-1}{T_c-T_1},
\end{displaymath}
which maximizes the likelihood. Since $\hat{\gamma}^c_{MLE}$ is asymptotically $\mathcal{N}(\gamma^c, \gamma^c/c)$ distributed \cite{FisherMLE}, one can form an asymptotic confidence interval for $\gamma^c$. Whenever $\hat{\gamma}^{c+1}_{MLE}$ is outside of the previous confidence interval, $T_c$ is deemed to be a change point. Here, the number of samples up until $T_c$ is $c$, which implies that the first $c$ sampling instants are distributed as the order statistics from $c$ uniformly distributed variables in $(T_1, T_c)$ \cite{AICiP}. The resulting mixture distribution, an example of which is shown in Fig. \ref{sampdisthist}, possess the desired characteristics associated to ice core data, with an example of the growing missampling errors being shown in Fig. \ref{missampvsT}.

Next, the global periodicities are simulated to mimic those that may be expected from the orbital theory of climate, having periods of $100,41,23$, and $19$ kyr \cite{dust,IAaAC,vostok}.
The base amplitudes $\rho_k$ for the frequencies are here set to be $1, 0.8, 0.6$, and $0.6$, with each data set having a deviation from this amplitude which is normal distributed with standard deviation $\rho_k/10$. The phase is drawn uniformly in $[0,\pi/5]$, in order to reflect the growth- and ablation rates of large ice sheets, which has a mean response time of roughly $10$ kyr \cite{growthRate}. It is worth noting that the method presented here takes into account the possibility of such phase shift, whereas the stacking method described below that is used for comparison does not. 

Furthermore, the used missampling is drawn from a normal distribution with a standard deviation adapted such that the probability that two samples switching place is less than $1\%$ (independently for each sample), automatically ensuring that the standard deviation of the missampling grows as the intensity function of the Poisson process decreases.

For each simulation, $3$ data sets are simulated, and the spectra for each are estimated using three different methods: the proposed estimator described above, the mean Lomb-Scargle periodogram \cite{LSper}, and the stacked periodogram \cite{BassinotStack,SPECMAP}, both the latter being typical examples of state-of-the-art techniques used in the field.

The mean periodogram is constructed by normalizing the data by $N_m$, such that each core contributes equally, and the mean of all periodograms is computed and considered to be the common spectrum. Since amplitude estimation accuracy is not considered in this application, the resulting estimate is scaled for ease of comparison in subsequent figures.

The stacked periodogram is constructed by normalizing all data, stacking them into a single time-series and calculating the Lomb-Scargle periodogram of the resulting, stacked, time series. This stacking method has been used frequently to recreate the global signal \cite{BassinotStack, ImbrieSPEC,Imbrie943}.

To compare the methods, the sum of the mean squared errors of the frequency estimates are presented as a function of $SNR$ (in dB). For white noise with variance $\sigma_\upsilon^2$, the $SNR$ is defined as
\begin{displaymath}
    SNR \triangleq 10\log_{10}\left(\frac{\sum_{k=1}^K (\rho_k^{(m)})^2}{2\sigma_\upsilon^2}\right).
\end{displaymath}

With a total of $N_R$ simulations of the $3$ data sets, let $\hat{f}^n_k$ denote the estimate of $f_k$ in the $n$:th simulation. The frequency $MSE$ is thus defined as 
\begin{displaymath}
    MSE_k \triangleq \frac{1}{N_R}\sum_{n = 1}^{N_R}(\hat{\omega}^n_k-\omega_k)^2.
\end{displaymath}
The frequency estimates for the periodogram-based methods are chosen as the $K$ largest maxima of the spectral estimate.

\subsection{Without missampling}\label{noMiss}
\begin{figure}[!t]
    \centering
    \includegraphics[width = .5\textwidth]{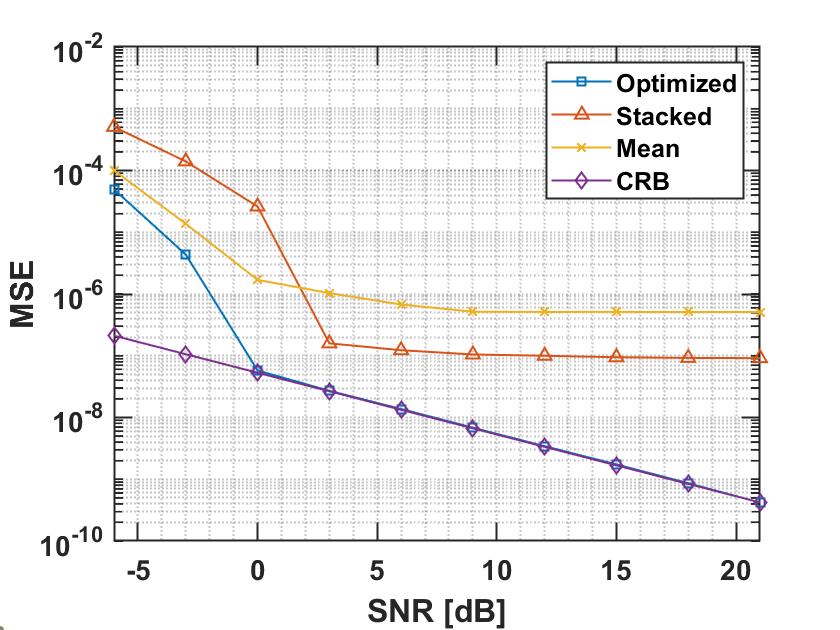}
    \caption{The sum of $MSE_k$ calculated from 1000 Monte-Carlo runs per $SNR$, as compared to the sum of averaged frequency CRB of the data assuming no missampling. From $SNR = 0$ dB, the proposed approach achieves the CRB, whereas the periodogram-based methods show limited reduction in $MSE$ for increasing $SNR$.}
    \label{MSE}
\end{figure}

\begin{figure}[!t]
    \centering
    \includegraphics[width = .5\textwidth]{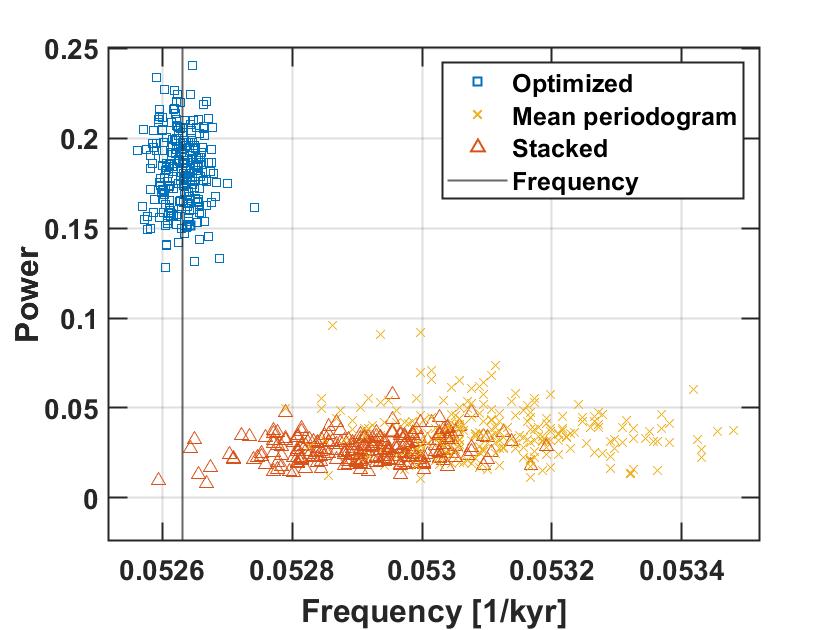}
    \caption{Frequency estimates of the highest frequency $\omega = 1/19$ kyr$^{-1}$ at $SNR = 15$ dB. The periodogram-based methods show significant bias due to the closeness of the peaks. The resulting sidelobes of the other frequency will as a result push the estimate to higher frequencies, creating a bias.}
    \label{freq_est}
\end{figure}

\begin{figure}[!t]
    \centering
    \includegraphics[width = .5\textwidth]{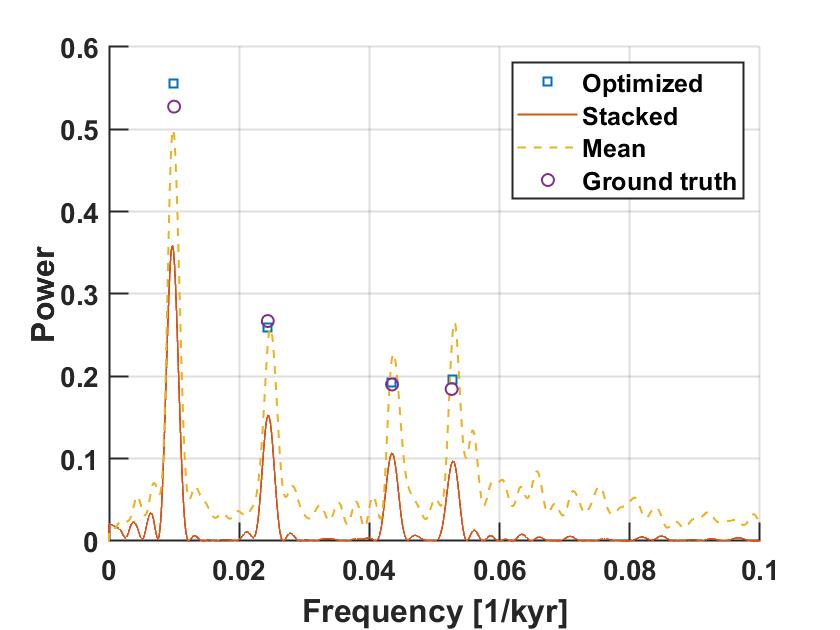}
    \caption{A typical estimate using the three methods as compared to the ground truth which is the mean power of the three data sets. The data is here simulated with $SNR = 5$ dB.}
    \label{Estimate}
\end{figure}

We initially examine the case when the sampling patterns for each data set is known. 
Fig. \ref{MSE} shows the sum of the frequency $MSE_k$ for the three methods when there is no missampling, as compared to the corresponding CRB. The results for the proposed method were obtained using 4 zooming steps, starting with 16 initital bins with active threshold $\tau = 10^{-5}$ $\lambda = 15$, $\zeta^{(m)} = 10$ for all $m$, and $\bq$ is a vector of ones.  
As is clear from the figure, the $MSE$ does not decrease for the periodogram-based methods. This effect is not due to a lack of decrease in variance as such, but rather due to bias of the frequency estimates, which can be seen in Figs. \ref{freq_est} and \ref{Estimate}. The proposed estimator on the other hand yields a statistically efficient estimate for $SNR$s above $0$ dB.

\subsection{With missampling}
\begin{figure}[!t]
    \centering
    \includegraphics[width = .5\textwidth]{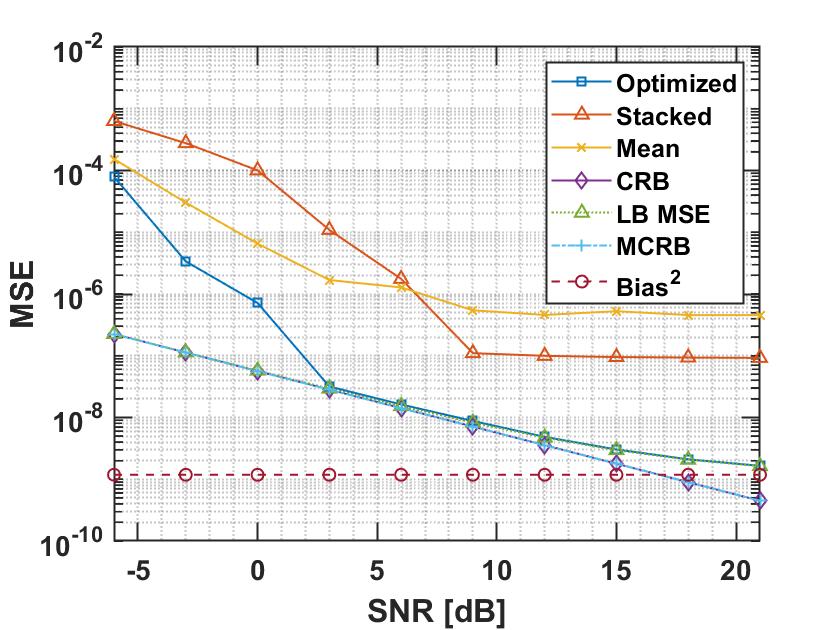}
    \caption{The sum of $MSE_k$ calculated from 2000 Monte-Carlo runs per $SNR$, as compared to the sum of averaged lower bound on the $MSE$ on the frequency estimates, the MCRB, squared bias of the pseudo-true frequencies, and the CRB assuming no missampling. The proposed approach achieves the lower bound on the $MSE$ from $SNR = 3$ dB. Above $SNR = 18$ dB, the squared bias is the largest contributing component of the misspecified $MSE$ lower bound.}
    \label{missampMSE}
\end{figure}
Fig. \ref{missampMSE} shows the results using the same setup as above, but now incorporating the expected missampling as described above. It is worth noting that the MCRB and the CRB yields almost the same bound, except for the higher $SNR$ cases, where the bias of the pseudo-true frequencies dominates such that the lower bound decreases more slowly. Since the missampling error does not change with the $SNR$, it seems reasonable that the difficulty of estimating the frequencies changes from initially being limited by the $SNR$ to eventually being limited by the missampling for higher $SNR$ cases. The same bias problems for the periodogram-based methods at higher $SNR$s remain, implying that the $MSE$ does not continue to decrease after $SNR = 9$ dB. Again, the proposed estimator yields a statistically efficient estimate from $SNR = 3$ dB. 

\subsection{Results using real data}\label{ResultsUsingRealData}
\begin{figure}[!t]
    \centering
    \includegraphics[width = .5\textwidth]{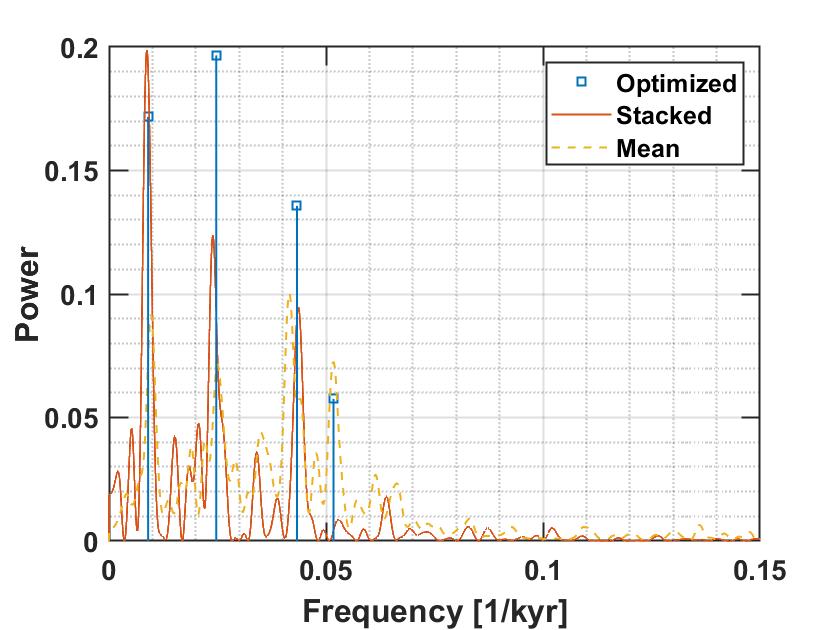}
    \caption{The estimated global spectrum using measured ice-core data from three different ice-cores. The resulting frequency estimates using the proposed method are $\omega_1 = 9.014\cdot 10^{-3}$, $\omega_2 = 2.471\cdot 10^{-2}$, $\omega_3 = 4.331\cdot10^{-2}$, and $\omega_4 = 5.173\cdot10^{-2}$ [kyr$^{-1}$] with periods of $110.9$, $40.47$, $23.09$, and $19.33$ kyr, respectively. Here, we used $\lambda = 15$, $\zeta = 10$, with $4$ zooming steps.}\label{realspect}
\end{figure}

Using data from three different ice cores, the Vostok ice core \cite{vostok}, Taylor Dome ice core \cite{taylorAge,taylorAge2,taylorDome}, and the Dome F ice core \cite{DOMEF}, the results of the method were computed; the results are shown in Fig. \ref{realspect} and the estimated frequencies can be seen to be fairly close to those expected from orbital theory with periods of $100, 41, 23$ and $19$ kyr \cite{IAaAC}. In Table \ref{resultTable}, the frequency estimates are listed according to which expected frequency they correspond to
\begin{table}[!t]\renewcommand{\arraystretch}{1.3}
\caption{Expected periods and corresponding estimated periods for each of the methods in kyr}
    \label{resultTable}
    \centering
    \begin{tabular}{|c|c|c|c|}
    \hline\hline
         Expected periods & Optimized & Stacked & Mean  \\
         \hline\hline
         100 & 110.9 & 114.61 & 103.04 \\
         \hline
         41 & 40.47 & 41.78 & 40.03\\
         \hline
         23 & 23.09 & 22.92 & 24.07\\
         \hline
         19 & 19.33 & Missing & 19.33\\
         \hline\hline
    \end{tabular}
\end{table}
Interestingly, the stacked approach only shows three of the expected peaks, while all four are visible in the resulting estimates using the other two approaches. Also worth noting is that the relative amplitudes for each estimate is different, such that when using the proposed estimation approach, the second peak with a period of $40.47$ kyr is the highest, whereas for the stacked method the first peak with a period of $114.61$ kyr is the highest, and for the last approach the third peak with a period of $24.07$ kyr is the highest. This shows that the choice of estimation approach significantly impacts the resulting estimate.

\section{Conclusions}\label{Conclusions}
In this work, we have introduced a sparse reconstruction technique to estimate the global spectrum corresponding to the shared spectral structures of a set of non-uniformly sampled data sets, each sampled at different times and with significant uncertainties in the sampling times. The method is shown to yield statistically efficient estimates, as well as reaching the corresponding misspecified performance bound. The performance of the method is shown using realistic simulations of ice core data, including the growing uncertainty in the sampling times as the age of the samples grow, as well as with measured ice core data.

\appendices
\section{A small example: Behaviour of the Schuster periodogram under missampling}\label{powerloss}
To exemplify the effects of missampling on the expected value of the periodogram, we consider and idealized example where the signal $s(t)$ consists of a single complex harmonic in circularly-symmetric white noise with total variance $\sigma_\upsilon^2$, such that
\begin{displaymath}
    s(t) = \rho e^{i\omega_1(t + \Delta_t)} + \upsilon_t
\end{displaymath}
which is then sampled at at equidistant sampling instants, $T$, where only the missampling $\Delta_t\sim \mathcal{N}(0, \sigma_\Delta^2) \ \forall t \in T$ is random, with the sampling times being evenly spaced. Then, the expected value of the periodogram at angular frequency $\omega$, can be expressed as
\begin{displaymath}
\begin{aligned}    
    \mathbb{E}\left\{\phi_h^{LS}(\omega)\right\} &=\mathbb{E}\left\{\frac{1}{N}\left|\sum_{t\in T} \left(\rho e^{i\omega_1(t + \Delta_t)}+\upsilon_t\right)e^{-i\omega t}\right|^2\right\} \\
    &=\frac{1}{N}\mathbb{E}\left\{\left|\sum_{t\in T}\rho e^{i((\omega_1-\omega)t+\omega_1\Delta_t)}+\upsilon_te^{-i\omega t}\right|^2\right\} \\
    &=\frac{1}{N}\mathbb{E}\left\{\left(\sum_{t \in T} \rho e^{i(\omega_1-\omega)t + i\omega_1\Delta_t}+\upsilon_te^{-i\omega t} \right)\right.\\ 
    &\left.\left(\sum_{n\in T}\rho e^{-i(\omega_1-\omega)n-i\omega_1\Delta_n} +\upsilon_n e^{i\omega n}\right)\right\} 
    \end{aligned}
\end{displaymath}
Thus,
\begin{displaymath}
\begin{aligned}
    \mathbb{E}\left\{\phi_s^{LS}(\omega)\right\} &=\frac{\rho^2}{N}\sum_{t\in T}\sum_{n \in T}\mathbb{E}\left\{e^{i(\omega_1-\omega)(t-n)+ i\omega_1(\Delta_t-\Delta_n)}\right\} \\
    &+\sum_{t\in T}\sum_{n \in T}\mathbb{E}\left\{\upsilon_n e^{i\omega n} \rho e^{i(\omega_1-\omega)t+i\omega_1\Delta_t}\right.\\&\left. + \upsilon_t e^{-i\omega t}\rho e^{-i(\omega_1-\omega)n-i\omega_1\Delta_n} + \upsilon_t\upsilon_n\right\} \\
    &=\frac{1}{N}\sum_{t\in T}\sum_{n \in T}\rho^2\mathbb{E}\left\{e^{i(\omega_1-\omega)(t-n)+ i\omega_1(\Delta_t-\Delta_n)}\right\}\\ &+ \mathbb{E}\{\upsilon_t \upsilon_n\}\\
    &= \frac{1}{N}\Big[\rho^2N + N\sigma_\upsilon^2\\
    &+\rho^2\sum_{t \in T}\sum_{n\in T: n \neq t}\mathbb{E}\left\{e^{i(\omega_1-\omega)(t-n)}e^{i\omega_1(\Delta_t-\Delta_n)} \right\} \Big]\\
    &= \frac{1}{N}\Big[N(\rho^2+\sigma_\upsilon^2) \\
    &+\rho^2\sum_{t \in T}\sum_{n \in T : n\neq t}e^{i(\omega_1-\omega)(t-n)}\mathbb{E}\{e^{i\omega_1(\Delta_t-\Delta_n)}\} \Big]
\end{aligned}
\end{displaymath}
where one can identify the characteristic function of $(\Delta_t-\Delta_n)\sim \mathcal{N}(0, 2\sigma_\Delta^2), \forall t,n\in T$, and thus
\begin{displaymath}
    \mathbb{E}\{e^{i\omega_1(\Delta_t-\Delta_n)}\} = e^{-\omega_1^2\sigma_\Delta^2}
\end{displaymath}
implying that
\begin{displaymath}
\begin{aligned}
    \mathbb{E}\left\{\phi_h^{LS}(\omega)\right\} &= \frac{1}{N}\Bigg[N(\rho^2+\sigma_\upsilon^2) \\
    &+ \left.2\rho^2\sum_{t\in T}\sum_{n\in T: n<t}\cos\left\{(\omega_1-\omega)(t-n)\right\}e^{-\omega_1^2\sigma_\Delta^2}\right]
\end{aligned}
\end{displaymath}
At $\omega = \omega_1$,
\begin{displaymath}
\begin{aligned}
    \mathbb{E}\{\phi_h^{LS}(\omega_1)\} &= \frac{1}{N}\left[N(\rho^2+\sigma_\upsilon^2) + 2\rho^2\sum_{t\in T}\sum_{n\in T:n<t} e^{-\omega_1^2\sigma_\Delta^2} \right]\\
    &=\frac{1}{N}\left\{N(\{\rho^2+\sigma_\upsilon^2)+\rho^2N(N-1)e^{-\omega_1^2\sigma_\Delta^2} \right\}\\
    &\leq \rho^2N+\sigma_\upsilon^2
\end{aligned}
\end{displaymath}
with equality if and only if $\sigma_\Delta^2 = 0$. Thus, given uncertainty in sampling times, power is lost.

\section{Calculating the KLD between two multivariate Gaussian normal distributions}\label{KLDapp}
Using the definition of the KLD and the true and approximating model PDFs as specified in \eqref{truemodel} and \eqref{mismodel}, respectively, one obtains 
\begin{displaymath}
\begin{aligned}
    D(f||g_{\btheta}) &= \mathbb{E}_{f}\left\{\log\left(\frac{f}{g}\right)\right\} \\
    &= \frac{N}{2}\log\left(\frac{\sigma_{\epsilon}^2}{\sigma_\upsilon^2}\right) +\frac{1}{2}\sum_{m=1}^M\sum_{t\in \bT^{(m)}}\Bigg[\frac{\left(\nu_t^{(m)}(\btheta)\right)^2}{\sigma_{\epsilon}^2}\\
    &\quad-\frac{\mathbb{E}_{f}\left\{\left(y_t^{(m)}-s_t^{(m)}\right)^2\right\}}{\sigma_\upsilon^2}\Bigg]\\ 
\end{aligned}
\end{displaymath}
thus, using the wave-form error $\kappa_t^{(m)}(\btheta) \triangleq s_t^{(m)}-\mu^{(m)}_t(\btheta)$
\begin{displaymath}
    \begin{aligned}
   D(f||g_{\btheta}) &= \frac{N}{2}\log\left(\frac{\sigma_{\epsilon}^2}{\sigma_\upsilon^2}\right)\\
   &\quad +\frac{1}{2}\sum_{m=1}^M\sum_{t\in \bT^{(m)}}\Bigg[\frac{\left(\kappa_t^{(m)}(\btheta)\right)^2}{\sigma_{\epsilon^2}} + \frac{\sigma_\upsilon^2}{\sigma_{\epsilon}^2}-1\Bigg] \\
    &= \frac{1}{2}\Bigg[N\log\left(\frac{\sigma_{\epsilon}^2}{\sigma_\upsilon^2}\right)+ N\left(\frac{\sigma_\upsilon^2}{\sigma_{\epsilon}^2}-1\right)\\
    &\quad +\frac{1}{\sigma_{\epsilon}^2}\sum_{m=1}^M\sum_{t\in\bT^{(m)}}\left(\kappa_t^{(m)}(\btheta)\right)^2\Bigg]
\end{aligned}
\end{displaymath}

\section{Calculating the two generalizations of the FIM}\label{FIMA}
With the true and approximating model PDFs as specified in \eqref{truemodel} and \eqref{mismodel}, respectively, one obtains
\begin{displaymath}
\begin{aligned}
    \bA_{\btheta_0} &= \mathbb{E}_{f}\left\{\nabla_{\btheta}^2 \ln{g_{\btheta}}(\by)\big|_{\btheta = \btheta_0}\right\}\nonumber \\
    &= \mathbb{E}_{f}\left\{\nabla_{\btheta}\frac{\nabla_{\btheta}^T g_{\btheta}(\by)}{g_{\btheta}(\by)}|_{\btheta=\btheta_0}\right\}\nonumber \\
    &= \mathbb{E}_{f}\left\{\nabla_{\btheta}^2\left[-\frac{1}{2\hat{\sigma}_{\epsilon}^2}\sum_{m=1}^M\sum_{t\in \bT^{(m)}}\left(\nu_t^{(m)}(\btheta_0)\right)^2\right]\right\}.
\end{aligned}
\end{displaymath}
Thus,
\begin{displaymath}
\begin{aligned}
    \bA_{\btheta_0} &= -\frac{1}{2\hat{\sigma}_{\epsilon}^2}\sum_{m=1}^M\sum_{t\in\bT^{(m)}}\mathbb{E}_{f}\left\{\nabla_{\btheta}^2\left(\nu_t^{(m)}(\btheta_0)\right)^2\right\}\nonumber \\
    &= \frac{1}{\hat{\sigma}_{\epsilon}^2}\sum_{m=1}^M\sum_{t\in \bT^{(m)}}\mathbb{E}_{f}\left\{(\nu_t^{(m)}(\btheta_0))\nabla_{\btheta}^2\mu_t^{(m)}(\btheta_0) \right.\nonumber\\
    &\quad-\left. \nabla_{\btheta}\mu_t^{(m)}(\btheta_0)\nabla_{\btheta}^T\mu_t^{(m)}(\btheta_0)\right\}\nonumber \\
    &= \frac{1}{\hat{\sigma}_{\epsilon}^2}\sum_{m=1}^M\sum_{t\in\bT^{(m)}}\left\{\left(\kappa_t^{(m)}(\btheta_0)\right)\nabla_{\btheta}^2\mu_t^{(m)}(\btheta_0)\right.\nonumber\\ 
    &\left.\quad-\nabla_{\btheta}\mu_t^{(m)}(\btheta_0)\nabla_{\btheta}^T\mu_t^{(m)}(\btheta_0)\right\},
\end{aligned}
\end{displaymath}
and 
\begin{displaymath}
    \begin{aligned}
        \bB_{\btheta_0} &= \mathbb{E}_{f}\left\{\nabla_{\btheta}\ln{g_{\btheta}}(\by)|_{\btheta = \btheta_0}\nabla_{\btheta}^T\ln{g_{\btheta}(\by)}|_{\btheta = \btheta_0}\right\}\\
        &=\frac{\sigma_\upsilon^2}{(\hat{\sigma}_{\epsilon}^2)^2}\sum_{m=1}^M\sum_{t\in \bT^{(m)}} \nabla_{\btheta}\mu_t^{(m)}(\btheta_0)\nabla_{\btheta}^T\mu_t^{(m)}(\btheta_0) \\
        &+ \frac{1}{(\hat{\sigma}_{\epsilon}^2)^2}\left[\bb_t^m(\bb_k^n)^T\right],
    \end{aligned}
\end{displaymath}
with $\bb_t^m = \sum_{m=1}^M\sum_{t\in\bT^{(m)}}\kappa_t^{(m)}(\btheta_0)\nabla_{\btheta}\mu_t^{(m)}(\btheta_0)$.
Since the pseudo-true parameters minimizes the squared $\ell_2$ norm between the approximating and true signal, it follows by first order optimality conditions that
\begin{displaymath}
\begin{aligned}
   \bb_t^m \propto \nabla_{\btheta}\left( \sum_{m=1}^M\sum_{t\in \bT^{(m)}}(\kappa_t^{(m)}(\btheta_0))^2 \right)  = \mathbf{0}.
\end{aligned}
\end{displaymath}
Thus, 
\begin{displaymath}
    \bB_{\btheta_0}= \frac{\sigma_\upsilon^2}{(\hat{\sigma}_{\epsilon}^2)^2}\sum_{m=1}^M\sum_{t\in \bT^{(m)}} \nabla_{\btheta}\mu_t^{(m)}(\btheta_0)\nabla_{\btheta}^T\mu_t^{(m)}(\btheta_0).
\end{displaymath}
It is worth noting that when the model is correctly specified, the MCRB simplifies to the CRB \cite{MCRB}.



\ifCLASSOPTIONcaptionsoff
  \newpage
\fi



\bibliographystyle{IEEEtran}
\bibliography{main_arxiv.bbl}
\end{document}